\documentclass[twocolumn,amstext,amsmath,amssymb,prl,preprintnumbers,superscriptaddress,floatfix,showpacs,nofootinbib]{revtex4}

\usepackage[dvips]{epsfig}
\usepackage{slashed}
\usepackage{mathrsfs}
\usepackage{graphicx}
\usepackage{xcolor}
\usepackage{hyperref}
\usepackage{dcolumn}
\usepackage{subfigure}
\usepackage{xspace}
\usepackage{ulem}

\newcommand{\GeV}{~\mathrm{GeV}}
\newcommand{\TeV}{~\mathrm{TeV}}

\newcommand{\s}{~\mathrm{s}}

\newcommand{\Equref}[1]{Eq.~(\ref{#1})}

\begin{document}
\preprint{LMU-ASC 11/16}
\preprint{TUM-HEP 1038/16}
\preprint{FLAVOUR(267104)-ERC-121}

\title{Dark matter decays from non-minimal coupling to gravity}

\author{Oscar Cat\`a}
 \affiliation{Ludwig-Maximilians-Universit\"at M\"unchen, 
   Fakult\"at f\"ur Physik,\\
   Arnold Sommerfeld Center for Theoretical Physics, 
   80333 M\"unchen, Germany}

  \author{Alejandro Ibarra}
\affiliation{Physik-Department, Technische Universit\"at M\"unchen, James-Franck-Stra\ss{}e, 85748 Garching, Germany}

  \author{Sebastian Ingenh\"utt}
\affiliation{Physik-Department, Technische Universit\"at M\"unchen, James-Franck-Stra\ss{}e, 85748 Garching, Germany}

\pacs{95.35.+d, 95.30.Cq}

\begin{abstract}

We consider the Standard Model extended with a dark matter particle in curved spacetime, motivated by the fact that the only current evidence for dark matter is through its gravitational interactions, and we investigate the impact on the dark matter stability of terms in the Lagrangian linear in the dark matter field and proportional to the Ricci scalar. We show that this ``gravity portal'' induces decay even if the dark matter particle only has gravitational interactions, and that the decay branching ratios into Standard Model particles only depend on one free parameter: the dark matter mass. We study in detail the case of a singlet scalar as dark matter candidate, which is assumed to be absolutely stable in flat spacetime due to a discrete $Z_2$ symmetry, but which may decay in curved spacetimes due to a $Z_2$-breaking non-minimal coupling to gravity. We calculate the dark matter decay widths and we set conservative limits on the non-minimal coupling parameter from experiments. The limits are very stringent and suggest that there must exist an additional mechanism protecting the singlet scalar from decaying via this gravity portal.
\end{abstract}

\maketitle

\section{Introduction}
\label{sec:intro}

Multiple astronomical and cosmological observations have by now demonstrated the existence of dark matter (DM), a non-luminous matter component in our Universe possibly constituted by a population of new particles not contained in the Standard Model (SM)  of Particle Physics, which were produced in the very first stages of our Universe by an as yet unidentified mechanism (for a review, see~\cite{Bertone:2004pz}). Within this framework the DM particle(s) are required to have a lifetime at least as large as the age of the Universe. Furthermore, if the DM particle decays producing photons, neutrinos or antimatter, the non-detection of excesses in the cosmic fluxes of these particles with respect to the expected backgrounds implies a DM lifetime orders of magnitude larger than the age of the Universe (for a review, see~\cite{Ibarra:2013cra}).

A simple possibility to ensure a sufficiently long lifetime consists in postulating a very light DM particle, such that the decay rate becomes naturally suppressed by the small phase space available. However, in many well motivated scenarios, notably those where the DM density is generated via thermal freeze-out, the DM particle is much heavier and a new symmetry, either global or local, must be introduced in order to ensure DM stability.

This argument is usually formulated in flat spacetime. However, since the only current evidence for DM is  through its gravitational effects, it seems natural to embed the DM framework in curved spacetime. This extension might lead to qualitative differences in the DM phenomenology. In particular, whereas gauge symmetries are expected to be exact even in curved backgrounds, this might not be the case for global symmetries. As a result, and unless the DM particle is protected against decay by a local symmetry, the Lagrangian may contain effective operators, suppressed by powers of the Planck mass, inducing DM decay into SM particles with a long, albeit potentially measurable, lifetime. 

In this paper we will investigate the possibility that the DM particle could decay via non-minimal coupling to gravity. We will assume that the DM is stable in flat spacetime due to a global symmetry, which is explicitly broken by gravitational interactions through a term proportional to the Ricci scalar, such that the global symmetry gets restored in the limit of flat spacetime. We will show that this ``gravity portal'' leads to DM decay, even if the DM does not have interactions with the SM, save for gravity. As an illustration, we will focus on the singlet scalar DM model, which is stabilized in flat spacetime by a global $Z_2$ symmetry, and we will explicitly calculate the decay channels and widths induced by a $Z_2$-breaking non-minimal coupling term to gravity. 

\section{Dark matter non-minimally coupled to gravity}
\label{sec:setup}

In curved spacetime, the classical action of the SM extended with a DM particle reads:
\begin{equation} \label{eq:CompleteActionJordan}
  \mathcal S = \int d^4x \sqrt{-g}\left[ -\frac{R }{2 \kappa^2}+ \mathcal L_\text{SM} + \mathcal L_\text{DM} \right]\;.
\end{equation}
Here $g$ is the determinant of the metric tensor $g_{\mu\nu}$, $R$ is the Ricci scalar, $\kappa=\bar M_P^{-1}=\sqrt{8 \pi G}$ is the inverse reduced Planck mass, $\mathcal L_\text{DM}$ is the part of the Lagrangian involving the DM particle and $\mathcal L_\text{SM}$ is the SM Lagrangian, which can be cast as:
\begin{align}
{\mathcal{L}}_\text{SM}&\equiv {\mathcal{T}}_{F}+{\mathcal{T}}_{f}+{\mathcal{T}}_{H}+{\mathcal{L}}_{Y}-{\mathcal{V}}_{H}\;,
\end{align}
where ${\cal{L}}_Y$ contains the Yukawa interactions, ${\mathcal{V}}_{H}$ is the Higgs potential and ${\mathcal{T}}_j$ are the kinetic terms of gauge bosons, fermions and scalars,
\begin{align}
{\mathcal{T}}_{F}&=-\frac{1}{4}g^{\mu\nu}g^{\lambda\rho}F_{\mu\lambda}^aF_{\nu\rho}^a\;,\nonumber\\
{\mathcal{T}}_{f}&=\frac{i}{2}{\bar{f}}\stackrel{\longleftrightarrow}{\slash{\!\!\!\!\nabla}}\!\!f\;,\nonumber\\
{\mathcal{T}}_{H}&=g^{\mu\nu}(D_{\mu}H)^{\dagger}(D_{\nu}H)\;.
\end{align}
Above we have defined $\slash{\!\!\!\!\nabla}=\gamma^ae^{\mu}_a\nabla_{\mu}$, $\gamma_c$ being a Dirac matrix, $e^{\mu c}$ a vierbein and ${\nabla}_\mu=D_\mu -\frac{i}{4}e_\nu^b (\partial_\mu e^{\nu c}) \sigma_{bc}$, with $D_\mu$ the gauge covariant derivative.

In this work we will assume that the DM particle is stable in flat spacetime but can decay through operators linear in the dark matter field containing the Ricci scalar, which are in general expected unless forbidden by a gauge symmetry. The lowest-dimensional operator of this class is linear in $R$ and can be generically written as 
\begin{equation} \label{eq:generalCoupling}
  \mathcal L_\xi = - \xi R F(\varphi,X)\;,
\end{equation}
with $\xi$ a dimensionless coupling parameter and $F(\varphi,X)$ a real function of the DM field $\varphi$ (and potentially other fields $X$), invariant under the Lorentz and SM gauge symmetries. Adding this non-minimal coupling term, the action reads
\begin{equation} \label{eq:CompleteActionJordanDM}
  \mathcal S = \int d^4x \sqrt{-g}\left[ -\frac{R }{2 \kappa^2}\Omega^2(\varphi,X) + \mathcal L_\text{SM} + \mathcal L_\text{DM} \right]\;,
\end{equation}
where 
\begin{equation}\label{eq:def-Omega}
\Omega^2(\varphi,X) =1 + 2 \kappa^2 \xi F(\varphi,X)\;.
\end{equation}

Eq.~(\ref{eq:CompleteActionJordanDM}) is written in the so-called Jordan frame, where the DM field is explicitly coupled to the curvature operator. Both quantities can be decoupled by performing a Weyl transformation on the metric tensor:
\begin{equation}
  \widetilde g_{\mu\nu} = \Omega^2(\varphi,X) g_{\mu\nu}\;,
\end{equation}
which leads to
\begin{equation} \label{eq:CompleteActionEinstein}
  \mathcal S=\int d^4x \sqrt{- \widetilde g}\left[ -\frac{\widetilde R}{2 \kappa^2} + \frac{3}{\kappa^2}\frac{\widetilde \nabla_\mu \Omega \widetilde \nabla^\mu \Omega}{\Omega^2} + \mathcal {\widetilde{L}}_\text{SM} +\mathcal {\widetilde{L}}_\text{DM} \right]\;,
\end{equation}
where tilded quantities refer to the transformed metric $\widetilde g_{\mu\nu}$. The action is now in the so-called Einstein frame, where the gravitational field equations take their canonical form. Explicitly, the Weyl-transformed SM Lagrangian reads:
\begin{align} \label{eq:SMLagrangianEinstein}
{\mathcal{\widetilde{L}}}_\text{SM}&={\mathcal{\widetilde T}}_{F}+\frac{1}{\Omega^{3}}{\mathcal{\widetilde T}}_{f}+\frac{1}{\Omega^2}{\mathcal{\widetilde T}}_{H}+\frac{1}{\Omega^4}({\mathcal{L}}_{Y}-{\mathcal{V}}_{H})\;, 
\end{align}
and contains, upon Taylor expanding $\Omega(\varphi,X)$ in the DM field $\varphi$, linear terms in the latter which in turn induce DM decay into SM particles.\footnote{Similar terms will also induce proton decay through this gravity portal. Its lifetime can however be estimated to be orders of magnitude larger than the present experimental bound.}

The concrete decay channels and rates depend on the nature of the DM particle. Nevertheless, from \Equref{eq:SMLagrangianEinstein} one can already identify some general features of DM decay through non-minimal coupling to gravity. First, the decay widths are suppressed by powers of the Planck mass and are proportional to the non-minimal coupling parameter $\xi$ squared. As a result, the decay branching rations will only depend on the DM mass. Second, since the gauge kinetic term remains invariant under the Weyl transformation, there is no fundamental coupling of the DM particle to diphotons or digluons. A coupling to the $Z$ and the $W$ boson pairs is, however, induced through the kinetic term of the Higgs boson. Third, in the absence of a cosmological constant term, there is no term in the Lagrangian inducing DM decay into gravitons at tree level.

In the remainder of the letter we will investigate in detail the gravitationally induced decays in the scalar singlet DM model and we will  discuss the observational constraints on the non-minimal coupling parameter $\xi$.

\section{Scalar singlet dark matter}
\label{sec:scalar_singlet}

Consider a real scalar field $\phi$, singlet under the SM gauge group and charged under a discrete $Z_2$ symmetry~\cite{Silveira:1985rk,McDonald:1993ex}. The part of the Lagrangian involving the DM particle reads, in the Jordan frame, 
\begin{equation}
  \mathcal L_\text{DM}=\frac{1}{2}g_{\mu\nu}\partial^\mu \phi \partial^\nu \phi -V(\phi,H)\;,
\end{equation}
where the interaction terms in the potential are assumed to be invariant under the transformation $\phi\rightarrow -\phi$. We assume that the $Z_2$ symmetry that ensures the stability of the DM holds in flat spacetime, but may be broken in curved spacetimes, leading to DM decay. The lowest-order $Z_2$-breaking term is 
\begin{equation} \label{eq:non-minimal-scalar-singlet}
  \mathcal L_\xi = - \xi M R \phi\;,
\end{equation}
where $M$ is a mass parameter introduced so that $\xi$ is dimensionless. For this interaction term, the Weyl factor defined in \Equref{eq:def-Omega} reads $\Omega^2(\phi)=1+ 2 \kappa^2 \xi M \phi$. Using \Equref{eq:SMLagrangianEinstein} and expanding $\mathcal {\widetilde{L}}_\text{SM} $ to first order in $\phi$ we obtain 
\begin{align} \label{eq:LSM_ScalarDM_Einstein}
{\mathcal{\widetilde{L}}}_\text{SM}&\supset -2\kappa^2\xi M\phi\left[\frac{3}{2}{\mathcal{\widetilde T}}_{f}+{\mathcal{\widetilde T}}_{H}+2({\mathcal{L}}_{Y}-{\mathcal{V}}_{H}) \right]\;,
\end{align}
which induce DM decay into SM particles.\footnote{Strictly, the field $\phi$ is not canonically normalized. However, the correction to the kinetic term due to non-minimal coupling is of order $\xi^2$ and therefore can be neglected when calculating the rate at the lowest order in $\xi$.} We note that this ``gravity portal'' induces decay even when the DM interaction terms with the Standard Model particles are absent. Besides, the higher dimensional operators in Eq.~(\ref{eq:LSM_ScalarDM_Einstein}) do not arise from integrating out heavy particles, as necessarily occurs in flat spacetime, but arise as a mere consequence of the non-minimal coupling to gravity when formulating the theory in the Einstein frame.

\begin{table}[tb]
\centering
\renewcommand{\arraystretch}{1.2}
  \begin{tabular}{| l | c |}
    \hline
    Decay mode &  Rate proportional to \\
    \hline
    $\phi\rightarrow hh, WW, ZZ$                          & $m_\phi^3$\\
    $\phi\rightarrow f\overline f$		                        & $m_f^2 m_\phi$\\
    \hline
    $\phi\rightarrow hhh$			                        & $v^2m_\phi$\\
    $\phi\rightarrow WWh, ZZh$	                        & $m_\phi^5/v^2$\\
    $\phi\rightarrow f\overline fh$		                        & $m_f^2 m_\phi^3/v^2$\\
    $\phi\rightarrow f\overline f'W, f\overline fZ$	& $m_\phi^5/v^2$\\
    $\phi\rightarrow f\overline f\gamma, q\overline qg$	& $m_\phi^3$\\
    \hline
    $\phi\rightarrow hhhh$			                        & $m_\phi^3$\\
    $\phi\rightarrow WWhh, ZZhh$			& $m_\phi^7/v^4$\\
    \hline
  \end{tabular}
  \caption{Tree level decay channels of the scalar singlet DM candidate induced by the non-minimal coupling to gravity, together with the parametric dependence of the decay rate on the DM mass and the SM mass scales.}
  \label{tab:phiDecays}
\end{table}

The different tree level DM decay channels are listed in table \ref{tab:phiDecays}, together with the dependence of the corresponding decay rates on the DM and SM masses. From Eq.~(\ref{eq:LSM_ScalarDM_Einstein}) it follows that all decay rates are suppressed by a common factor $\xi^2 M^2/ \bar M_P^{4}$. Hence, on dimensional grounds, the decay rate must contain an additional factor with dimensions of mass to the third power. The precise form of this factor is listed for each channel in the right column of table \ref{tab:phiDecays}. 

For DM masses below the electroweak scale, the dominant decay channel is expected to be  $\phi\to q{\bar{q}}g$ (or $\phi\to f {\bar{f}} \gamma$ when the hadronic decays are kinematically forbidden), due to the helicity suppression of the rate for $\phi\to f{\bar{f}}$. On the other hand, for very large DM masses the four-body decays $\phi\to WWhh,ZZhh$ are expected to dominate, due to the enhancement in the rate by a factor $m_{\phi}^4/v^4$. Lastly, for intermediate masses the three-body decays $\phi\to f{\bar{f}}'W,f{\bar{f}}Z$ are expected to dominate, except in the rather small window $v\lesssim m_{\phi}\lesssim 4\pi v$, where two-body decays with high thresholds, namely $\phi\to WW,~ZZ,~hh,~t{\bar{t}}$ have rates of comparable size to the phase-space suppressed three-body channels. 

\begin{figure}[t]
\begin{center}
\includegraphics[width=0.45\textwidth]{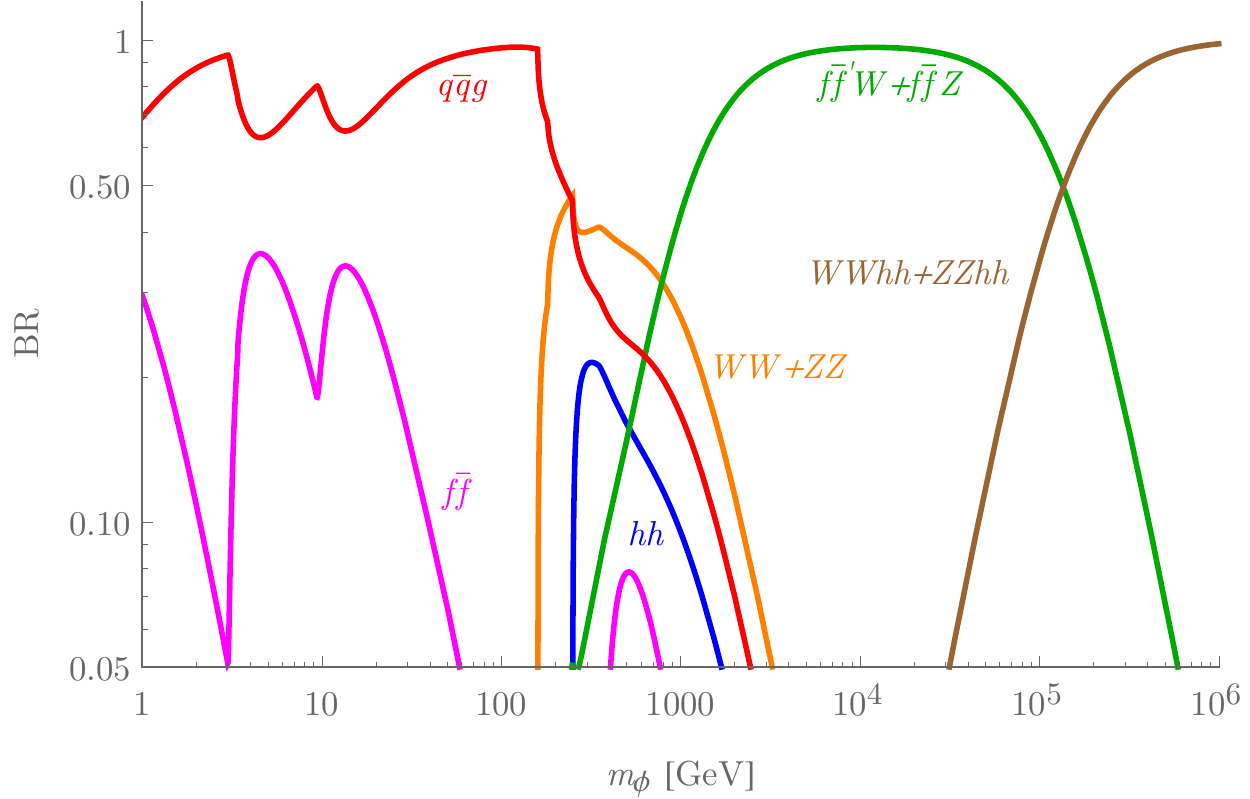} 
\end{center}
\caption{\small Decay branching fractions of the scalar singlet DM via non-minimal coupling to gravity.}
\label{fig:BRs} 
\end{figure}

The exact values of the branching ratios are shown in Fig.~\ref{fig:BRs}; the decay processes $\phi\to f\bar{f}\gamma$, $\phi\to 3h$, $\phi\to 4h$, $\phi\to f\bar{f}h$ and $\phi\to WWh,~ZZh$ have branching ratios below 5\% and are not shown in the Figure.  As anticipated above, the dominant decay modes are $\phi\to q{\bar{q}}g$ for $m_\phi \sim 1-200\,\text{GeV}$, $\phi\to WW,~ZZ,~hh,~t{\bar{t}},~q{\bar{q}}g$ for $m_\phi \sim 200\,\text{GeV}- 1 \TeV$, $\phi\to f{\bar{f}}'W,~f{\bar{f}}Z$ for $m_\phi \sim 1- 100  \TeV$ and $\phi\to WWhh,~ZZhh$ for $m_\phi\gtrsim 100  \TeV$. Details about the calculation will be presented in a forthcoming publication~\cite{CII:2016}.

The requirement that the DM particle has a lifetime longer than the age of the universe and, more importantly, the non-observation of the gamma-rays, antimatter particles and neutrinos produced in the decay over the expected astrophysical backgrounds, sets strong limits on the non-minimal coupling $\xi$. For the two-body decay channels $\phi\rightarrow WW,~ZZ,hh, f\bar f$ detailed studies exist. In particular, limits on the partial width into $WW$ have been derived in \cite{Cirelli:2012ut} employing the extragalactic gamma-ray flux extracted from the Fermi-LAT data \cite{Ackermann:2014usa}, and read $\Gamma^{-1}_{WW}\gtrsim 10^{27}\s$ for $m_\phi\sim 0.2 \TeV- 30 \TeV$. Complementary limits have been obtained using the AMS-02 positron data \cite{Ibarra:2013zia} and antiproton data \cite{Giesen:2015ufa}. In the $m_\phi\sim 0.2 \TeV- 30 \TeV$ mass range, the limits on the decay width from neutrino data are weaker \cite{Covi:2009xn}. However, for heavier DM particles, neutrino experiments provide the most stringent bounds on the (inverse) decay width: for decay channels producing hard neutrinos, the limit on the inverse width is $\Gamma^{-1}\gtrsim 10^{26}-10^{27}\,\text{s}$ for $m_\phi\sim 10 \TeV-10^{15} \GeV$~\cite{Esmaili:2012us}. 

A dedicated analysis of the gravitationally induced three- and four-body decays in the scalar singlet DM scenario is beyond the scope of this work. In view of the bounds resulting from neutrino experiments, we will impose the conservative lower limit $\Gamma^{-1}\gtrsim 10^{24}\,\text{s}$ over the whole mass range for all the decay channels.

\begin{figure}[t]
\begin{center}
\includegraphics[width=0.45\textwidth]{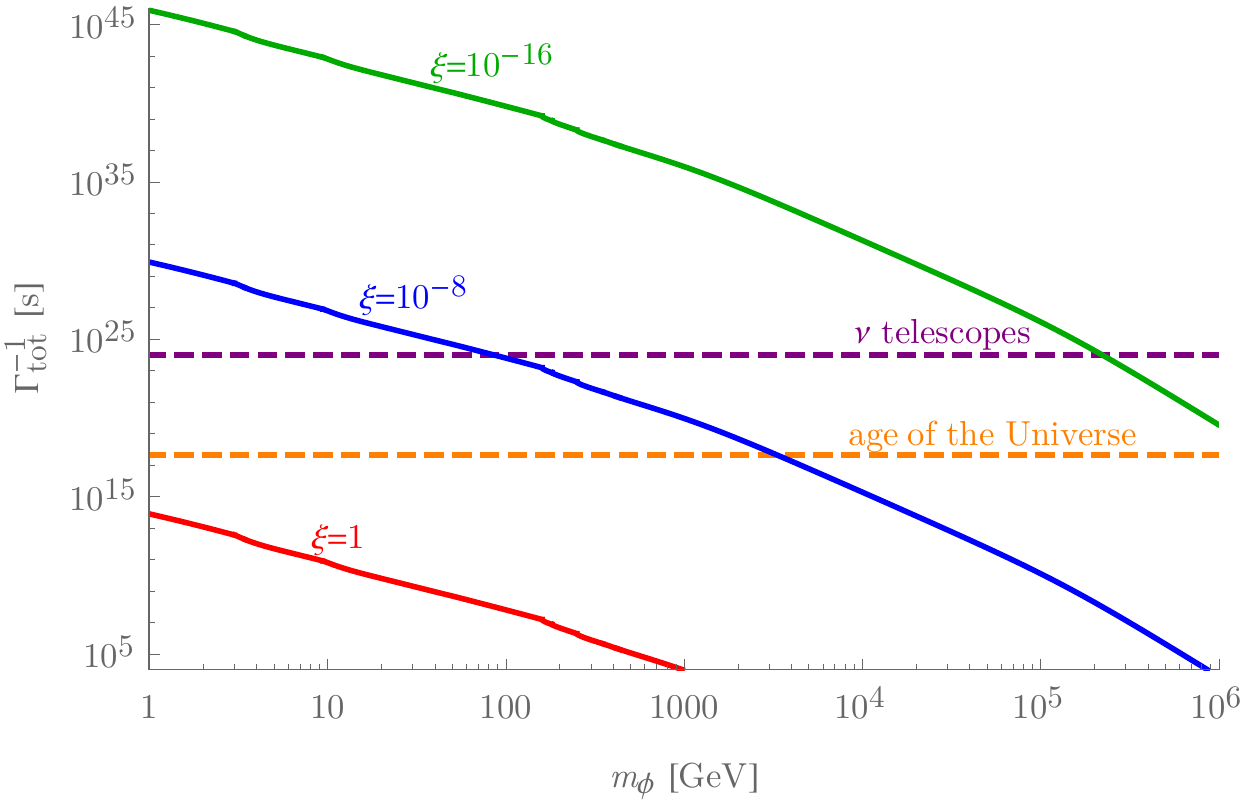} 
\end{center}
\caption{\small Inverse total decay width as a function of the DM mass, assuming $M=\bar M_P$ and for $\xi=1,~10^{-8},~10^{-16}$. For comparison, the age of the Universe and an approximate lower limit on the inverse decay width from neutrino telescopes are also shown.}
\label{fig:inverse_width} 
\end{figure}

We show in Fig.~\ref{fig:inverse_width} the inverse total decay width of the scalar singlet DM for non-minimal coupling parameters $\xi=1,~10^{-8}$ and $10^{-16}$, assuming $M=\bar M_P$. We also show for comparison the age of the Universe, $\tau_U\approx 4 \times 10^{17}\,\text{s}$, and the conservative lower limit on the inverse decay width from neutrino telescopes $\Gamma^{-1}\gtrsim 10^{24}\,\text{s}$. As apparent from the plot, non-minimal coupling parameters ${\cal O}(1)$ lead to a DM lifetime which is orders of magnitude shorter than the age of the Universe. Furthermore, the non-observation in experiments of the neutrinos produced in the decay require $\xi\lesssim 10^{-8}~(10^{-16})$ for $m_\phi\sim 100 \GeV~(3\times 10^5  \GeV)$.
 
More concretely, an approximate lower bound on the total decay width is
\begin{align}
\displaystyle{
\Gamma_\text{tot}\gtrsim \frac{\xi^2}{8\pi} \frac{M^2 m_\phi^3}{\bar M_P^4} \times \begin{cases}
 \displaystyle 2n_q\,\frac{\alpha_s}{\pi}, & m_\phi\sim 1-200  \GeV,   \\
\displaystyle{1+2n_q\,\frac{\alpha_s}{\pi}},& m_\phi\sim 0.2-1 \TeV,  \\
\displaystyle\frac{3}{(2\pi)^2}\frac{m_\phi^2}{v^2},& m_\phi\sim 1-100  \TeV, \\
\displaystyle\frac{1}{10(8\pi)^4}\frac{m_\phi^4}{v^4}, & m_\phi\gtrsim 100  \TeV,
 \end{cases}}
\label{eq:total-Gamma-limit}
\end{align}
which follows from considering just the dominant decay channel in the given mass range,  where $n_q$ is the number of quarks kinematically accessible in the decay. Therefore, the requirement $\Gamma^{-1}_\text{tot}\gtrsim 10^{24}\,\text{s}$ translates into
\begin{align}
\left|\frac{\xi M}{{\bar{M}}_P}\right| \lesssim  \begin{cases}
 2\times10^{-8}\,\left(\frac{m_\phi}{100 \GeV}\right)^{-3/2},  & m_\phi\sim 1-200  \GeV,   \\
8\times10^{-10}\,\left(\frac{m_\phi}{500 \GeV}\right)^{-3/2},& m_\phi\sim 0.2-1  \TeV,  \\
2\times10^{-14}\,\left(\frac{m_\phi}{50 \TeV}\right)^{-5/2},&m_\phi\sim 1-100  \TeV, \\
4\times10^{-15}\,\left(\frac{m_\phi}{100 \TeV}\right)^{-7/2},& m_\phi\gtrsim 100  \TeV.
 \end{cases}
\label{eq:xi-limit}
\end{align}
Clearly, for the singlet scalar DM model, there must exist a mechanism suppressing the non-minimal coupling to gravity, especially for very large DM masses.

\section{Conclusions}

We have investigated the possibility that the dark matter could decay via a non-minimal coupling to gravity. The leading operator of this class is linear in the dark matter field and in the Ricci scalar, and is in general present in the Lagrangian, unless forbidden by a gauge symmetry. The decay rates through this ``gravity portal'' are suppressed by powers of the Planck mass and are proportional to the square of the non-minimal coupling parameter. Notably, the rates are non-vanishing  even if the dark matter only has gravitational interactions. Furthermore, the  decay branching ratios into Standard Model particles depend only on one free parameter, the dark matter mass, rendering a fairly predictive framework. 

We have illustrated dark matter decays via this ``gravity portal'' in a simple scenario, where the dark matter particle is a scalar singlet which is stabilized in flat spacetime by a $Z_2$ symmetry. We have assumed that the non-minimal coupling to gravity breaks the $Z_2$ symmetry and we have identified the dominant decay channels. Finally, we have calculated a constraint on the non-minimal coupling parameter from the non-observation in experiments of an excess in the cosmic neutrino fluxes. The limits are very stringent and strongly indicate that an additional stabilization mechanism must be at work in the scalar singlet dark matter scenario, possibly based on local invariance, in order to ensure a sufficiently long-lived dark matter particle.

\vspace{0.5cm}
{\it Acknowledgements}: We thank Gia Dvali, Jos\'e Ram\'on Espinosa, C\'esar G\'omez and Andi Trautner for discussions. This work was partially supported by the DFG cluster of excellence ``Origin and Structure of the Universe'' and by the ERC Advanced Grant project ``FLAVOUR''(267104).


\bibliography{references}

\end{document}